# Electric-field-induced Three-terminal pMTJ Switching in the absence of an External Magnetic Field

Jiefang Deng,[1,2,a)] Xuanyao Fong,[1] and Gengchiau Liang[1,2,b)]

[1]*Department of Electrical and Computer Engineering, National University of Singapore, Singapore 117576*

[2]*Centre for Advanced 2D Materials and Graphene Research Centre, National University of Singapore, Singapore 117546*

**Abstract**

Since it is undesirable to require an external magnetic field for on-chip memory applications, we investigate the use of a Rashba effective field alternatively for assisting the electric-field-induced switching operation of a three terminal perpendicular magnetic tunnel junction (pMTJ). By conducting macro-spin simulation, we show that a pMTJ with thermal stability of 61 can be switched in 0.5 ns consuming a switching energy of 6 fJ, and the voltage operation margin can be improved to 0.8 ns. Furthermore, the results also demonstrate that a heavy metal system that can provide large field-like torque rather than damping-like torque is favored for the switching.

Spin transfer torque (STT) based magnetic random access memory (MRAM) has been extensively studied because of its potential to improve current memory systems[1-3]. However, the STT-MRAM[4], whose cell structure consists of magnetic tunnel junctions (MTJ) and access transistors, requires a current density larger than $10^{10}$ A/m$^2$ to write the bit cell, which inevitably results in large write energy (~100 fJ)[5-7]. To reduce the switching current, the switching time, unfortunately, has to be sacrificed[4]. Moreover, for a two terminal STT-MTJ, since writing and reading processes share the same path, read disturb problem becomes severe. Therefore, it is difficult to optimize the writing and the reading processes independently[8-11].

Voltage control of magnetic anisotropy (VCMA) has been proposed as a solution to reduce the switching energy[12,13]. First principle studies demonstrate that an electric field can change the 3d-orbitals occupancies at the insulator/ferromagnet interface via spin-orbit interaction (SOI)[14-16], which modifies the magnetic anisotropy in the ferromagnet. Recently, experimental studies have demonstrated MTJs switched by VCMA[12,17-19], among which the minimum switching energy is as low as 6 fJ with switching times of 0.5 ns[18]. Despite the attractive improvements like low energy consumption and fast switching in previously reported VCMA based MTJs, the demand for an external magnetic field to assist the switching makes it unfeasible for practical memory applications. Our previous work has used an elliptical perpendicular MTJ (pMTJ) with a self-generated bias field to eliminate the requirement of an external magnetic field source[20], whereas the ever-present bias field sharply reduces the thermal stability to 28.

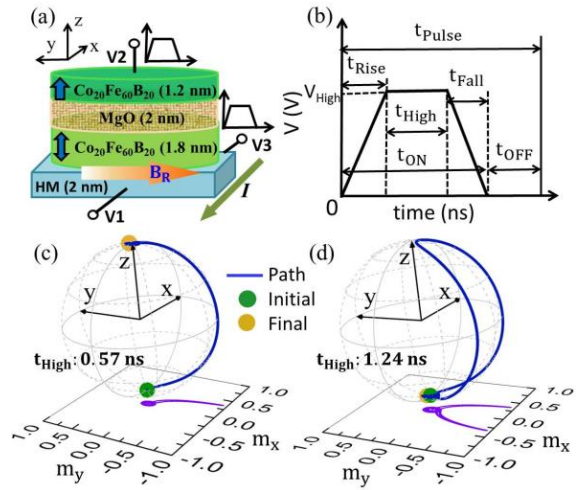

FIG. 1. (a) Schematic of a three-terminal elliptical pMTJ with a cross-section of 40 nm × 150 nm (major axis along y-axis). No external field is required for the VCMA-based switching. (b) Applied voltage pulse for V2 and V3 with $t_{Rise}$ and $t_{Fall}$ of 50ps. V1 is connected to the ground. The FL switching dynamics based on magnetization precession process for voltage pulses with $t_{High}$ ($t_{ON}$) of (c) 0.57 ns (0.67 ns) and (d) 1.24 ns (1.34 ns). The half precession period $t_{Half}$ is 0.67 ns in this case. For voltage pulses with $t_{ON}$ in odd multiples of $t_{Half}$, the FL toggles. Purple curves are the projections of the switching trajectories in x-y plane.

In this work, we study a device structure with an elliptical pMTJ on a heavy metal (HM) to overcome the aforementioned limitations. The Rashba effective field, **B**$_R$, generated at the ferromagnet/HM interface assists the VCMA to switch the pMTJ. Since **B**$_R$ is constrained to the MTJ being switched, the thermal stability of idle MTJs can maintain as high as 61. Moreover, this three-terminal structure separates the writing and the reading current paths, which mitigates the read disturb problem in STT-MTJs and improve the reading performance[21]. By comprehensively exploiting magnetic and electronic effects on the device performance, we demonstrate that in the absence of an external magnetic field, the pMTJ can be switched in 0.5 ns with small energy consumption of 6 fJ.

The studied pMTJ with a cross-section of 40 nm × 150 nm on a 2 nm thick HM is shown in Fig. 1(a). The MTJ major-axis is in y direction. A pinned-layer (PL, top $Co_{20}Fe_{60}B_{20}$) and a free-layer (FL, bottom $Co_{20}Fe_{60}B_{20}$) sandwich a 2 nm insulator layer (MgO). Both the PL and the FL have perpendicular easy axis. Unipolar voltage pulses in Fig. 1(b) with amplitudes V2$_{High}$ and V3$_{High}$ are applied to terminals V2 and V3, respectively, while V1 is grounded during the switching process. Positive V2$_{High}$ and V3$_{High}$ generate currents along respective $-\hat{z}$ and $-\hat{x}$. When positive V2$_{High}$ is applied to perform an electric field *E* across the MTJ, the perpendicular interfacial anisotropy K$_I$ in the FL is reduced, and its easy axis changes from the z-axis to the y-axis. Simultaneously, positive V3$_{High}$ is applied, and **B**$_R$ pointing in $-\hat{y}$ is generated at the FL/HM interface via Rashba effect, assisting the FL magnetization **M**$_{FL}$ (its unit magnetization is **m**) in precessing around the newly formed easy axis direction, i.e. the y-axis. By controlling the voltage pulse duration, t$_{ON}$, half-cycle precession or full-cycle precession can be obtained as shown in Fig. 1(c) and (d), correspondingly. After the voltages are removed, the FL easy axis recovers to the z-axis, and **M**$_{FL}$ relaxes to either $-\hat{z}$ or $\hat{z}$ depending on whether its final state is below or above the x-y plane.

Magnetization dynamics of the FL are modeled using the Landau-Lifshitz-Gilbert (LLG) equation[22] with the macro-spin approximation, which is valid for the FL dimensions considered in this work,

$$\frac{\partial \mathbf{m}}{\partial t} = -\gamma \mathbf{m} \times (\mu_0 \mathbf{H_{Eff}} + \mathbf{B_R}) + \alpha \mathbf{m} \times \frac{\partial \mathbf{m}}{\partial t} + \Gamma_{MTJ\_DL} \mathbf{m} \times \mathbf{p_{PL}} \times \mathbf{m} + \Gamma_{MTJ\_FL} \mathbf{m} \times \mathbf{p_{PL}} + \eta_{SOT} \mathbf{m} \times \mathbf{B_R} \times \mathbf{m}. \quad (1)$$

The first two terms on the right hand side are respective precession term and damping term, where $\gamma$, $\mu_0$ and $\alpha$ are the gyromagnetic ratio, the vacuum permeability and the damping constant, accordingly. The effective field $\mathbf{H_{Eff}}$ is contributed by magnetic anisotropy field, demagnetizing field[23,24] and thermal fluctuation[25]. The Rashba effective field is determined by[26]

$$\mathbf{B_R} = \frac{0.5\, \alpha_R}{\mu_B\, M_S} (\hat{\mathbf{z}} \times \mathbf{J_{FL}}), \quad (2)$$

where $\alpha_R$ is the Rashba coefficient, $\mu_B$ is the Bohr magneton, $\mathbf{J_{FL}}$ is the current density through the FL with a cross-section in y-z plane. In addition to the Rashba field, the damping-like torque (DLT) contributed by the SOI is also considered as shown in the last term. The third term and the fourth term in Eq. (1) are the respective STT DLT and field-like torque (FLT). For an interested reader, the detailed calculations of $\mathbf{H_{Eff}}$ and STT can refer to Ref. [20]. Parameters used in this work are summarized in Table I.

TABLE I. Parameters used in the simulation.

| Symbol | Parameter | quantity |
| --- | --- | --- |
| $M_S$ | Saturation magnetization | $1.26 \times 10^5$ A/m[27] |
| $K_{I0}$ | Zero-bias interfacial anisotropy | $1.3 \times 10^{-3}$ J/m$^2$[27] |
| $K_{U\_Bulk}$ | Bulk anisotropy | $2.24 \times 10^5$ J/m$^3$[27] |
| $\xi$ | VCMA coefficient | 50 fJ/(V·m)[13] |
| $TMR_0$ | Zero-bias tunnel magnetoresistance ratio | 144%[19] |
| $V_{Half}$ | Voltage at TMR = 0.5 $TMR_0$ | 0.45 V[19] |
| RA | MTJ resistance-area product | 1820 Ω·μm$^2$[19] |
| $\varepsilon_{MgO}$ | Relative permittivity of MgO | 9.7 |
| $\rho_{HM}$ | Resistivity of heavy metal layer | $2 \times 10^{-6}$ Ω·m[28] |

The FL switching time depends on the precession period, which is inversely proportional with the effective field. Therefore, V3$_{High}$, which controls **B$_R$**, would affect the switching process. Figure 2 illustrates the phase diagram of the MTJ switching probabilities from parallel (P) to anti-parallel (AP) and from AP-to-P (P10 and P01, respectively) *vs*. V3$_{High}$ and t$_{High}$. Each switching probability is calculated by conducting 100 LLG simulations under identical conditions. The red (dark blue) region indicates that the switching probability is 1 (0), which also refers to operation (retention) window. Pure VCMA induced switching is symmetric for P-to-AP and AP-to-P processes. It can be found that P10 in Fig. 2(a) and P01 in Fig. 2(b) exhibit almost the same trend, indicating that the STT, which always attempts to switch the MTJ in AP state since electrons always flow from the FL to the PL, is negligible. At a fixed V3$_{High}$, P10 or P01 oscillates between 0 and 1 as t$_{High}$ increases, consistent with the precession-based switching mechanism. For increasing t$_{High}$, the oscillating red or blue regions shrink because the magnetization is precessing and attending to relax along the y-axis, and then the precession period reduces. When t$_{High}$ is large enough, the switching probability would approach 0.5 because the large t$_{High}$ allows **M$_{FL}$** to fully relax along the y-axis. Once the voltage is removed, **M$_{FL}$** recovers to either $\hat{z}$ or $-\hat{z}$ in the presence of thermal fluctuations. In addition, for V3$_{High}$ of 2 mV, an operation window as large as 0.8 ns is obtained, which decreases as V3$_{High}$ increases. This is due to the fact that the increase of **B$_R$** increases the effective field along the y-axis, resulting in the reduction of the precession period, and then the decrease of operation window. It is noteworthy that the red regions indicate deterministic switching whereas the dark blue region can be designed for reading, which would greatly reduce the read disturb failure.

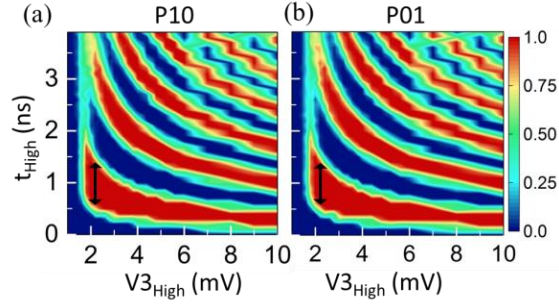

FIG. 2. Phase diagrams of switching probability from (a) P-to-AP (P10) and (b) AP-to-P (P01) as a function of $t_{High}$ and $V3_{High}$. Each switching probability is calculated using 100 simulations under identical conditions and with thermal fluctuations. Red region indicates a switching probability of 1 and thus the operation window, which has a maximum 0.8 ns for $V3_{High}$ of 2.2 mV.

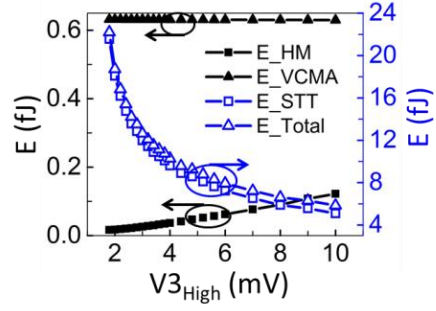

FIG. 3. Energy consumption including the Joule heating by the MTJ (E_STT) and the HM (E_HM), charging energy by the MTJ (E_VCMA) and the total energy (E_Total), as a function of $V3_{High}$. A minimum total energy of 6 fJ is achieved for $V3_{High}$ of 10 mV.

The total energy consumption and its three components are shown in Fig. 3. Interestingly, the total energy consumption decreases and reaches a minimum of 6 fJ in this work as $V3_{High}$ is increased. This is because increasing $V3_{High}$ would reduce the voltage drop across the MTJ $V_{MTJ}$ for a constant $V2_{High}$, and then reduce the Joule heating by the MTJ, which is proportional to the square of $V_{MTJ}$. As exhibited in Fig. 3, this Joule heating is more than one order larger than the other two energy components. In consequence, its reduction would dominate the reduction of the total energy consumption.

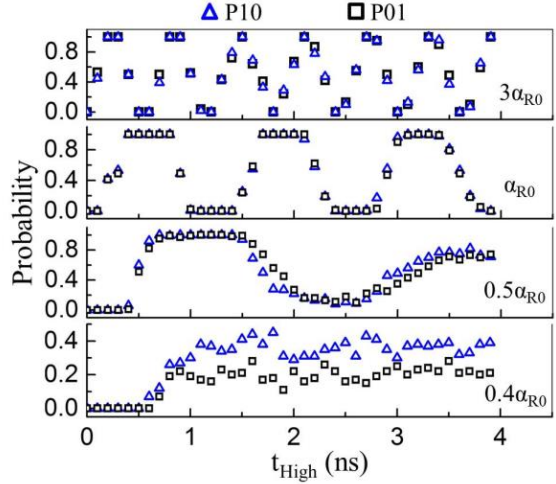

FIG. 4. Switching probabilities P10 and P01 as a function of $t_{High}$ with different Rashba coefficients $\alpha_R$. The $\alpha_{R0}$ is $10^{-10}$ eV·m. $V2_{High}$ and $V3_{High}$ are fixed to 2.5 V and 4 mV, respectively. The minimum $\alpha_R$ required for deterministic switching is $0.5 \times 10^{-10}$ eV·m, corresponding to an effective field of 13 mT per $10^{11}$ A/m².

In addition to $V3_{High}$, the Rashba coefficient $\alpha_R$, also determines the magnitude of $\mathbf{B_R}$ and thus the switching process. We then studied the switching probabilities P10 and P01 as a function of $\alpha_R$ for $V2_{High} = 2.5$ V and $V3_{High} = 4$ mV to explore the minimum $\alpha_R$ required for deterministic switching as displayed in Fig. 4. The $\alpha_{R0}$ is $10^{-10}$ eV-m, extracted from Ref. [26]. It is shown that similar oscillation behavior is also observed as $t_{High}$ increases, and the number of oscillation cycles reduces when $\alpha_R$ reduces. At $\alpha_R = 0.4\alpha_{R0}$, the oscillation behavior disappears, and the switching probability approaches 0.4 or 0.2, implying that no deterministic switching is obtained. This can be explained that the overly reduced $\mathbf{B_R}$ cannot support the magnetization to precess across the x-y plane. Due to thermal fluctuations, the magnetization would only flip to its opposite z-direction with small probability. Moreover, there is a difference between P10 and P01 especially at $\alpha_R$ of $0.4 \times 10^{-10}$ eV·m. This asymmetry is contributed to the STT effect. Since the STT effect favors AP state, it would increases the probability of flipping to AP state. As a result, P10 is slightly larger than P01. More importantly, Fig. 4 also demonstrates that the minimum required $\alpha_R$ to have deterministic switching is $0.5 \times 10^{-10}$ eV·m, corresponding to a $\mathbf{B_R}$ or SOI-induced FLT of 13 mT per $10^{11}$ A/m$^2$. This $\alpha_R$ can be further reduced to $0.2 \times 10^{-10}$ eV·m, i.e. FLT of 5.3 mT per $10^{11}$ A/m$^2$, when the V3$_{High}$ is increased to 10 mV. Despite the fact that a large FLT of 23 mT per $10^{11}$ A/m$^2$ is achieved experimentally for Ta/CoFeB/MgO structure[29], and even larger FLT like 100 mT per $10^{11}$ A/m$^2$ [26] or 29 mT per $10^{11}$ A/m$^2$ [30] can be obtained for Pt/Co/AlO system, further explorations of large FLT considering the CoFeB thickness engineering are demanded for the MTJ switched by an electric field.

It has been demonstrated that there are both FLT and DLT contributed by SOI in a system with a ferromagnetic layer on a HM layer[31-34]. In this work, we consider that the FLT is from Rashba effect with $\alpha_R$ to be $10^{-10}$ eV·m[26], and explore the required range of the DLT for deterministic switching. The magnitude of DLT is considered by varying the coefficient $\eta_{SOT}$, which is the magnitude ratio of DLT to FLT. Figure 5 shows the switching probabilities versus $\eta_{SOT}$ and $t_{High}$. To have deterministic switching, $\eta_{SOT}$ should be between -1 and 1.5. If the DLT has an opposite effect to the FLT, i.e. $\eta_{SOT}$ is negative, the precession process around the y-axis would be suppressed. Further increasing the magnitude of negative DLT would destroy the precession around the y-axis, and lead to a switching probability less than 1 in the presence of thermal fluctuations. On the contrary, if $\eta_{SOT}$ is positive, the DLT would hasten the precession, and make the FL magnetization to align along the y-axis more easily. Therefore, the deterministic operation window shrinks as the DLT further increases. Hence, the HM that can provide larger FLT rather than DLT in a HM/CoFeB/MgO system such as that in Ref. [29,35] would be favored for the VCMA switching in our device structure.

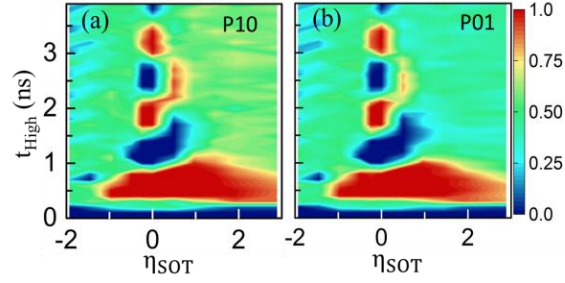

FIG. 5. Phase diagrams of switching probability from (a) P-to-AP and (b) AP-to-P as a function of $t_{High}$ and the damping-like torque (DLT). The $\eta_{SOT}$ is the ratio of DLT to field-like torque (FLT). To have a deterministic switching, $\eta_{SOT}$ should be in the range of -1 to 1.5.

In conclusion, we investigate a three-terminal pMTJ that is switched by the VCMA effect. The Rashba effective field generated at the HM/FL interface works as a bias field to assist the precession-based switching process, and thereby, no external magnetic field is required. This improvement on one hand enhances the FL thermal stability to 61, on the other hand makes the device more feasible for applications. Through macro-spin simulation, it is found that the pMTJ can be switched as fast as 0.5 ns with a write energy as low as 6 fJ. Besides, the operation window can be improved to 0.8 ns. To achieve deterministic switching, the required SOI FLT can be as small as 5.3 mT per $10^{11}$ A/m$^2$. These results indicate that our proposed device is promising for MRAM applications. Moreover, we also show that materials together with CoFeB/MgO generating large FLT rather than DLT would be attractive and demanded for the VCMA precession-based devices.

**Corresponding Author**

a) dengjiefang@u.nus.edu, b) elelg@nus.edu.sg

## ACKNOWLEDGEMENTS

This work at the National University of Singapore was supported by CRP award no. NRF-CRP12-2013-01 and MOE2017-T2-1-114. The authors gratefully acknowledge the funding from the National Research Foundation, Prime Minister Office, Singapore, under its Medium Sized Centre Programme.